\documentclass[12pt,preprint]{aastex}

\begin{document}

\title{A First Estimate Of The X-Ray Binary Frequency As A Function Of
  Star Cluster Mass In A Single Galactic System}

\author{D.~M. Clark \altaffilmark{1,2}, S.~S. Eikenberry \altaffilmark{2},
B.~R. Brandl \altaffilmark{3}, J.~C. Wilson \altaffilmark{6},
J.~C. Carson \altaffilmark{5}, C.~P. Henderson \altaffilmark{4},
T. ~L. Hayward \altaffilmark{7}, D.~J.  Barry \altaffilmark{4},
A.~F. Ptak \altaffilmark{8}, E.~J.~M. Colbert \altaffilmark{8}}

\altaffiltext{1} {Instituto de Astronom\'{i}a, Universidad Nacional
  Aut\'{o}noma de M\'{e}xico, Apdo Postal 877, Ensenada, Baja
  California, M\'{e}xico; dmclark@astrosen.unam.mx}
\altaffiltext{2}{Department of Astronomy, University of Florida,
  Gainesville, FL 32611; dmclark@astro.ufl.edu}
\altaffiltext{3}{Leiden University, P.O. Box 9513, 2300 RA Leiden,
  Netherlands.}
\altaffiltext{4}{Astronomy Department, Cornell University, Ithaca, NY
  14853.}
\altaffiltext{5}{Max Planck Institute for Astronomy, K\"{o}nigstuhl 17, D-69117 Heidelberg, Germany}
\altaffiltext{6}{Department of Astronomy, P.O Box 400325, University
  of Virginia, Charlottesville, VA 22904.}
\altaffiltext{7}{Gemini Observatory, AURA/Casilla 603, La Serena,
  Chile}
\altaffiltext{8}{Department of Physics and Astronomy, Johns
  Hopkins University, 3400 North Charles St., Baltimore, MD 21218.}

\begin{abstract}

We use the previously-identified 15 infrared star-cluster counterparts
to X-ray point sources in the interacting galaxies NGC 4038/4039 (the
Antennae) to study the relationship between total cluster mass and
X-ray binary number.  This significant population of X-Ray/IR
associations allows us to perform, for the first time, a statistical
study of X-ray point sources and their environments.  We define a
quantity, $\eta$, relating the fraction of X-ray sources per unit mass
as a function of cluster mass in the Antennae.  We compute cluster
mass by fitting spectral evolutionary models to $K_s$ luminosity.
Considering that this method depends on cluster age, we use four
different age distributions to explore the effects of cluster age on
the value of $\eta$ and find it varies by less than a factor of four.
We find a mean value of $\eta$ for these different distributions of
$\eta$ = 1.7$\times$10$^{-8}$ $M_{\sun}^{-1}$ with $\sigma_{\eta}$ =
1.2$\times$10$^{-8}$ $M_{\sun}^{-1}$.  Performing a $\chi^2$ test, we
demonstrate $\eta$ could exhibit a positive slope, but that it depends
on the assumed distribution in cluster ages.  While the estimated
uncertainties in $\eta$ are factors of a few, we believe this is the
first estimate made of this quantity to ``order of magnitude''
accuracy.  We also compare our findings to theoretical models of open
and globular cluster evolution, incorporating the X-ray binary
fraction per cluster.

\end{abstract}

\keywords{galaxies: starburst -- galaxies: star clusters -- X-rays:
binaries}

\section{Introduction}

The Antennae are a pair of colliding galaxies with an unusually large
number of X-ray point sources.  High resolution X-ray observations
taken with {\it Chandra} revealed 49 new individual sources
\citep{fab01}, where previous observations only indicated extended
filamentary structure \citep{fab97}.  These X-ray sources range in
luminosity from $10^{38}-10^{40}$ ergs s$^{-1}$.  Most of these are
thought to be X-ray binaries (XRBs) with a black hole compact
companion \citep{fab01}.

In addition to numerous X-ray sources, the Antennae contain many
bright, massive young clusters that are evident in both optical, {\it
HST} \citep{whi99} and infrared (IR) \citep[henceforth Paper I]{bra05}
images.  This makes this pair of interacting galaxies an ideal target
for studying the environments of XRBs.  In \citet[henceforth Paper
II]{cla07} we performed an extensive study of the XRB environments
using {\it Chandra} X-ray images and $J$ and $K_s$ IR images
(Paper I).  Our present paper will expand on our previous study by
exploring the relationship between XRBs and cluster mass in the
Antennae.

Recent theoretical models of young, massive cluster evolution provide
a framework for comparison to our observational study.  Two in
particular, \citep{osk05,sep05}, incorporate the fraction of XRBs per
cluster in their models.  \citet{osk05} use a population synthesis
code to study the evolution of X-ray emission in young, massive
clusters.  \citet{sep05} investigate the role of supernova kicks in
XRB expulsion from the parent cluster using the population synthesis
code, StarTrack.  They also incorporate the number of XRBs for a range
in cluster mass.  We will compare our measurements of the fraction of
XRBs per cluster in the Antennae to those predicted by these models.

We organize our paper as follows: In \S2 we give a brief summary of
our previous work on the Antennae and then define a quantity, $\eta$,
relating the XRB fraction to cluster mass in the Antennae.  In our
analysis, we estimate cluster mass using $K_s$ luminosity, which
depends non-trivially on the assumed cluster age.  We explore cluster
age/luminosity relations and their impact on our mass estimates in
\S3.  We compare $\eta$ to the measured value predicted by theoretical
cluster evolutionary models and present conclusions in \S4.

\section{Observations and Data Analysis}

\subsection{Infrared Images}

This paper is based on infrared (IR) $J$ (1.25 $\mu$m) and $K_s$ (2.15
$\mu$m) images of the Antennae galaxies.  We initially presented these
data and discussed the details of their reduction in Paper I.  In
summary, 20-minute total exposures in each filter were acquired using
the Wide-field InfraRed Camera (WIRC -- see \citet{wil03} for details)
on the Palomar 5-m telescope during the night of March 22nd, 2002.  In
Paper II, we made a frame-tie between the IR and X-ray images using IR
counterparts to circumvent the poor absolute astrometric accuracy of
{\it Chandra} ($\sim1\farcs5$).  We matched seven IR sources from the
WIRC images with {\it Chandra} X-ray point sources.  Using a least
squares fit of a linear matching function we tied {\it Chandra} right
ascension and declination to WIRC {\it x, y} pixel positions.  The rms
positional uncertainty is $\sim0\farcs5$.  With a strong astrometric
frame-tie in place we were able to accurately identify IR counterparts
to X-ray sources.  We found 19 IR counterparts within $1\farcs5$ of an
X-ray source, 13 of which were within 1$\farcs$0 of an X-ray source.
After estimating the IR source density, we predict only two of
the ``strong'' matches (separations $<1\farcs0$) and three of the
``possible'' matches (separations between 1$\farcs$0 -- 1$\farcs$5)
are due to chance superpositions of unrelated objects.

In Paper II, we pointed out two important implications for these
results.  First, that there is clearly a significant excess of IR
counterparts within $1\farcs0$ of the X-ray sources -- 13, where we
expect only two in the null hypothesis of no physical counterparts.
Even including the ``possible'' counterparts out to $1\farcs5$, we
have a total of 19 counterparts, where we expect only five are chance
superpositions.  Secondly, this implies that for any given ``strong''
IR counterpart, we have a probability of $\sim$ 85\% ($11/13$ with a
$1\sigma$ uncertainty of 0.3\footnote{Found using confidence levels
  for small number statistics listed in Tables 1 and 2 of
  \citet{geh86}.}) that the association with an X-ray source is real.
Even for the ``possible'' counterparts, the probability of true
association is $\sim$50\%.  Therefore, regardless of the physical
separations between the X-ray sources and their IR counterpart, we are
confident that the majority of these associations are real.

We note that of the 19 X-ray sources with counterparts, two are the
nuclei \citep{zez02a}, one is a background quasar \citep{cla05}, and
two share the same IR counterpart.  Therefore, in this paper we will
only consider the 15 IR counterparts (of the original 19) that are
star clusters in the Antennae.

\subsection{Photometry}

We performed aperture photometry in both the $J$ and $K_s$ bands on
all 15 IR cluster counterparts plus an additional 204 clusters
identified by eye in these IR images of the Antennae (see also Paper I
and the tables there-in).  We defined our aperture as $\sim3\sigma$ of
the Gaussian PSF, where $J$ had a full width at half maximum (FWHM) of
$1\farcs2$ and $K_s$ had a FWHM of $0\farcs9$.  We measured a mean and
median sky background flux in two separate annuli between $\sim$6 --
10$\sigma$ of the PSF.  To account for the exceptionally crowded field
of the Antennae, we employed the use of background arcs instead of
annuli. Multiplying these four measurements by the area of the central
aperture and then subtracting these from the flux in the central
aperture yielded four separate source flux measurements.  We defined
error in sky background, $\sigma_{sky}$, as the standard deviation of
the four measured source fluxes.  We also considered Poisson noise,
$\sigma_{adu}$, defined as the total source flux divided by the square
root of the gain for the WIRC instrument.  The gain for WIRC during
the observations was $2e^{-} DN^{-1}$ \citep{wil03}\footnote{At the
time of the Antennae observations, WIRC was equipped with a Hawaii-1
1K$\times$1K detector and this is the gain for it.}.  Adding
$\sigma_{sky}$ and $\sigma_{adu}$ in quadrature, we computed the total
error in flux, $\sigma_{flux}$.  We converted fluxes to magnitudes
using a bright, 2MASS star in the field and defined the error in
magnitude, $\sigma_m$, as $\sigma_{flux}$ divided by the mean flux.
Typical errors in magnitude were $\sim$0.06 mag in both bands, with no
error above 1.0 mag.

To estimate cluster masses we needed to compute $K_s$ luminosity
($M_{K_s}$).  We computed $M_{K_s}$ using reddening derived from $(J -
K_s)$ colors (Paper II).  Assuming all clusters are dominated by O
and B stars, their intrinsic $(J-K_s)$ colors are $\sim$0.2 mag.
Approximating this value as 0 mag, this allowed us to estimate
$A_{K_s}$ as $\simeq$ $(J-K_s)_{obs}$/1.33 using the extinction law
defined in \citet{car89}.

\subsection{XRB-to-Cluster Mass Fraction}

We assume for now that cluster mass is proportional to $K_s$
luminosity -- i.e. that the stellar composition of all clusters is the
same. We defined a luminosity cutoff for statistical purposes as
$M_{K_s}$ = $-13.2$ mag (see Paper II for details).  We binned the
data by 0.2 mag in $M_{K_s}$ and then calculated an average flux per
bin.  Computing the fraction of the total number of clusters per
average flux of each bin, we took this as the probability of finding a
cluster with a specific mass.  In Figure 1 we compare this probability
for both clusters with X-ray sources and all clusters in the Antennae.
This shows that XRBs are more common in more massive clusters.

This result is not surprising -- as star cluster mass increases, so
does the number of massive stars in it.  Through stellar evolution, a
certain fraction of these stars will die in supernova explosions,
leaving behind neutron star or black hole remnants.  In turn, a
fraction of these stellar remnants will retain/acquire a mass-donating
companion star, becoming a detectable XRB.  Thus, through sheer
numbers of stars in more massive clusters, we expect a greater
likelihood of finding XRBs in them.  This leads us to two interesting
questions: 1) quantitatively, what cluster mass will more likely
produce an XRB and 2) is there some intrinsic property of massive
cluster physics that favors the production of XRBs, beyond simple
scaling with mass?

We believe that our large sample of IR-to-X-ray associations provides
the first dataset sufficient to estimate the answers to these
questions for the Antennae galaxies.  In our approach to answer these
questions we explore the relationship between the number of X-ray
detections per unit mass as a function of cluster mass in the
Antennae.  We can formalize this expression in the following equation:

\begin{equation}
N_X(M_c) = N_{Cl}(M_c)\cdot\eta(M_c)\cdot M_c
\end{equation}

Here, $N_X(M_c)$ is the number of detected X-ray sources with an IR
cluster counterpart, $N_{Cl}(M_c)$ is the number of detected clusters,
and $\eta(M_c)$ is the fraction of X-ray sources per unit mass, all as
a function of cluster mass, $M_c$.

If $\eta(M_c)$ increases or decreases over a range in $M_c$, this
means there could be something peculiar about massive cluster physics
to favor or suppress XRB formation.  In contrast, a constant
$\eta(M_c)$ across all $M_c$ would indicate that more massive clusters
are more likely to have an XRB simply because they have more stars.

While $\eta(M_c)$ is a powerful tool in studying the number of XRBs
per cluster, it requires that we know the mass of each star
cluster. However, extrapolating the masses of the Antennae clusters
from our photometric data required models that called for estimates of
ages and metallicities.  While we successfully constrained these
inputs and determined cluster masses (see below and \S3), we first
sought to compute $\eta(M_c)$ in terms of a purely observable quantity
-- flux. Calculating $\eta(M_c)$ as a function of $K_s$-band flux,
$\eta(F_{K_s})$, allowed us to investigate non-model dependent trends
in $\eta(M_c)$.

We calculated $\eta(F_{K_s})$ for clusters with a $K_s$-band
luminosity brighter than the -13.2 mag cutoff using bin sizes of
$F_{K_s} = 4\times10^6$ in counts (DN) (Figure 2). This bin size was
small enough to show a trend in $\eta(F_{K_s})$, but large enough to
contain at least two clusters with X-ray sources, allowing us to
assign error bars to each value of $\eta(F_{K_s})$.  The errors
plotted on the graph are the measurement uncertainty in the mean value
of the four $\eta(F_{K_s})$ added in quadrature with the Poisson
uncertainty of the mean $\eta(F_{K_s})$ in each bin.  Due to the small
sample size per bin, we computed these uncertainties using the small
number statistics formulae described in \citet{kee62}.  Figure 2 shows
that $\eta(F_{K_s})$ is roughly consistent with a constant value of
$5.4\times10^{-8}$ $F_{K_s}^{-1}$ with an uncertainty of
$\sigma_{\overline{\eta}}= 1.8\times10^{-8}$ $F_{K_s}^{-1}$.

In essence, $\eta(F_{K_s})$ is comparing two different mass
distributions, $N_{Cl}(F_{K_s})$, the mass distribution for all
clusters in the Antennae and $N_X(F_{K_s})/F_{K_s}$, the mass
distribution for clusters with X-ray sources, normalized by flux.  If
there is a constant number of X-ray sources per unit cluster mass as
suggested by Figure 2, then these two mass distributions should be the
same.  We can further corroborate this result by comparing
$N_X(F_{K_s})/F_{K_s}$ and $N_{Cl}(F_{K_s})$ using a two-sided
Kolmogorov-Smirnov (K-S) test.  The K-S test yielded a D-statistic of
0.75 and a probability of 0.107 that they are related.  Considering
the separate cluster mass populations as two probability
distributions, each can be expressed as a cumulative distribution.
The D-statistic is then the absolute value of the maximum difference
between each cumulative distribution.  This test quantitatively
demonstrates that there is nothing peculiar about massive clusters in
the Antennae with associated XRBs.  We also computed the Pearson {\it
  r} linear correlation coefficient between $N_X(F_{K_s})/F_{K_s}$ and
$N_{Cl}(F_{K_s})$, finding a value of 0.99.  Since a value of 1 means
a perfect linear fit, this value of {\it r} further substantiates the
observed relationship in $\eta(F_{K_s})$.

We then converted $\eta(F_{K_s})$ into the more conventional units of
solar mass.  Here we assume all clusters are coeval.  Selecting
$M_{K_s}$ listed in the Bruzual-Charlot (BC) cluster evolutionary
models \citep{bru03} for a 20 Myr, $1M_{\sun}$ cluster as a typical
value in the Antennae \citep{whi99}, we converted the model $M_{K_s}$
to $F_{K_s}$ using the standard relationship between luminosity and
flux.  Multiplying $\eta(F_{K_s})$ by $F_{K_s}$ we converted
$\eta(F_{K_s})$ to solar masses: assuming a cluster metallicity of z =
0.02, $\eta = 5.8\times10^{-8} M_{\sun}^{-1}$ with an uncertainty of
$\sigma_{\overline{\eta}} = 1.9\times10^{-8}M_{\sun}^{-1}$, while
assuming a metallicity of z = 0.05, $\eta =
8.9\times10^{-8}M_{\sun}^{-1}$ with an uncertainty of
$\sigma_{\overline{\eta}} = 3.0\times10^{-8}M_{\sun}^{-1}$.

While we assume all clusters are $\sim$20 Myr old, we note that the
actual range in ages should be $\sim$1--100 Myr \citep{whi99}.  The BC
models indicate that clusters in this age range could vary by a factor
of as much as 100 in mass for a given $K_s$ luminosity.  Since $\eta$
is a function of mass, incorrectly assigning cluster ages has the
potential to significantly impact $\eta$.  In the next section, we
explore how differences in cluster age can affect the value of $\eta$.

\section{Effects of Age and Slope in $\eta$}

We investigated the effect differences in cluster age has on $\eta$ by
assuming three individual age distributions for the Antennae clusters:
an instant burst in which all clusters are the same age, a uniform
distribution, and a distribution of the form, $dN/d\tau \propto
\tau^{-1}$ \citep{fal05}.  In each case, we assumed all clusters have
solar metallicity (z = 0.02) \citep{whi99}.

In addition, we address the issue of whether the functional form of
$\eta(M_c)$ can be fit by a slope or if, indeed, $\eta(M_c)$ is
consistent with a single value, by performing a $\chi^2$ test.  We
computed the $\Sigma\chi^2$ between the $\eta$ values for each bin and
the mean value of $\eta$, and computed the $\Sigma\chi^2$ between the
$\eta$ values for each bin and a fitted line to these values.  The
difference between the two value of $\Sigma\chi^2$,
$\Delta\Sigma\chi^2$, indicated the significance of the fitted
slope. If $\Delta\Sigma\chi^2$ was less than one, with in one
$\Sigma\chi^2$ deviation, then this indicated that the fitted slope
was insignificant.  While a value of $\Delta\Sigma\chi^2$ between one
and two, with in two $\Sigma\chi^2$ deviations, indicated only a weak
slope.  Any value of $\Delta\Sigma\chi^2$ greater than two meant a
significant, nonzero slope in $\eta$.  For our initial case, where we
estimated cluster mass using $F_{K_s}$ we found $\Delta\Sigma\chi^2 =
1.9$, suggesting a weak, but non-negligible slope, in $\eta$.

Considering the case of an instant burst in star formation, we
assigned the same age to describe all clusters and picked several such
values in the range 1 -- 100 Myr.  Applying the BC models, we
converted $\eta(F_{K_s})$ to units of solar mass for a range in ages.
For each distribution in $\eta$ we computed a mean value (Figure 3).
The mean for these values is $\eta_{instant}$ = 3.3$\times$10$^{-8}$
$M_{\sun}^{-1}$ with a standard deviation of $\sigma_{instant}$ =
2.9$\times$10$^{-8}$ $M_{\sun}^{-1}$.

Performing our $\chi^2$ test for each distribution in $\eta$ (see
Figure 3), we found a mean in $\Delta\Sigma\chi^2$ of 0.86, with a
standard deviation, $\sigma_{\Delta\Sigma\chi^2}$ = 1.7.  This seems
to indicated conflicting results.  For some cases in assumed cluster
age there is no significant slope in $\eta$, while other cases show a
pronounced slope.  This becomes more evident by examining Figure 3,
which shows a range in $\Delta\Sigma\chi^2$ of $\sim$0--6.5.
Incidentally, for $\eta_{20}$, $\Delta\Sigma\chi^2$ = 2.2, while for
$\eta_{100}$, $\Delta\Sigma\chi^2$ = 4.8$\times$10$^{-3}$.

Next, we assumed a uniform age distribution for our Antennae cluster
sample.  Picking ages at random from a uniform distribution between 1
-- 100 Myr, we assigned an age to each cluster in our sample.
Applying the BC models, we computed each cluster's mass based on the
assigned age, produced a mass distribution, and then calculated a mean
$\eta$.  Performing a Monte Carlo (MC) simulation, we recreated
cluster mass distributions 10,000 times, producing a large sample of
$\eta$'s with a mean of $\overline{\eta}_{uniform}$ = 5.4$\times$10$^{-9}$
$M_{\sun}^{-1}$ and $\sigma_{\overline{\eta}_{uniform}}$ = 6.3$\times$10$^{-10}$
$M_{\sun}^{-1}$ (Figure 4).

We then administered our $\chi^2$ test for each realized distribution
of $\eta$, finding a mean $\overline{\Delta\Sigma}\chi^2$ = 5.9 and
$\sigma_{\overline{\Delta\Sigma}\chi^2}$ = 5.2.  In some cases, $\eta$
didn't exhibit a positive slope.  Therefore, we investigated whether
the slope in $\eta$ tended to be mostly positive or mostly negative by
multiplying each $\overline{\Delta\Sigma}\chi^2$ statistic by the sign
of the slope.  In doing so, we found a mean
$\overline{\Delta\Sigma}\chi^2$ = 5.8 and
$\sigma_{\overline{\Delta\Sigma}\chi^2}$ = 5.3.  Thus almost all
distributions in $\eta$ for the uniform case had a positive slope.

A more realistic approach is assuming the cluster ages are defined by
a power law (PL): $dN/d\tau \propto \tau^{-1}$ \citep{fal05}.  These
authors derived their relationship using {\it HST} {\it
  UBVI}H$_{\alpha}$ observations of $\sim$11,000 clusters.  Fitting BC
models to photometry of each cluster, they generated an age
distribution.  In our analysis we picked ages at random according to
this distribution.  Mirroring the procedure used for the uniform
distribution case above, we created a sample of 10,000 $\eta$ values.
We found a mean for this sample of $\overline{\eta}_{PL}$ =
8.7$\times$10$^{-9}$ $M_{\sun}^{-1}$ and a
$\sigma_{\overline{\eta}_{PL}}$ = 1.2$\times$10$^{-9}$
$M_{\sun}^{-1}$.

Performing our $\chi^2$ test as we did with the uniform age
distribution case, we found a mean $\overline{\Delta\Sigma}\chi^2$ =
5.1 and $\sigma_{\overline{\Delta\Sigma}\chi^2}$ = 6.0.  Accounting
for variations in the sign of the slope in $\eta$, we found
$\overline{\Delta\Sigma}\chi^2$ = 4.3 and
$\sigma_{\overline{\Delta\Sigma}\chi^2}$ = 6.6.  Therefore, the trend
in the slope of $\eta$ remains positive, but not as significantly as
the uniform age distribution case.

In a final scenario, we used the cluster ages listed in the
electronic table available through \citet{men05} to fit ages to 144
clusters in our sample, including all 15 clusters associated with
X-ray sources.  \citet{men05} derived ages by using three age
indicators -- $UBVI$ and $K_s$ broadband photometry to break the
age/reddening degeneracy, H$\alpha$ and Br$\gamma$ emission to
identify clusters less than 7 Myr, and CO band-head absorption from
narrow-band images for clusters $\sim$10 Myr.  They then fit these
data to theoretical spectra for ages $<500$ Myr using a $\chi^2$
minimization technique \citep[for details, see][]{men05}.

Following the method discussed in \S2.3, we used the BC models to
convert $M_{K_s}$ to mass for those clusters with \citet{men05} age
estimates.  Using cluster bins of 1.0$\times10^7M_{\sun}$ in mass, we
computed four values for $\eta$ (see Figure 5).  The errors plotted on
the graph are uncertainties in the mean value of $\eta$ added in
quadrature with the Poisson uncertainty in each bin.  Again, we used
the small number statistic formulae in \citet{kee62} to compute these
errors.  We found a mean in $\eta$ of 2.2$\times10^{-8}M_{\sun}^{-1}$
and $\sigma_{\overline{\eta}} = 1.2\times10^{-8}M_{\sun}^{-1}$.
Applying our $\chi^2$ test, we found $\Delta\Sigma\chi^2$ =
1.0$\times$10$^{-2}$, implying no significant slope in $\eta$.

We summarize our age analysis in Figure 6.  Comparing the
distributions in $\eta$ for the fitted ages, instant burst, uniform
and power law age distributions, all values for $\eta$ are within a
factor of four.  Assuming an instant burst of 20 Myr, $\eta$ differs
by a factor of ten (see Figure 6).  Excluding specific ages for instant
bursts of cluster formation, there is little variation in the value of
$\eta$.

Comparing our $\chi^2$ test for each of the four different age
assumptions indicates inconsistent results (see Table 1).  In the
instant burst, uniform and power law cases, the slope in $\eta$ varies
from insignificant to distinctly positive, while $\eta$ has no
significant slope when the ages derived by \citet{men05} are fit to
our clusters.  Therefore, we can not ignore that $\eta$ might have a
non-zero slope and we will discuss the implications of this in the
following section.

\section{Summary and Conclusions}

Through the quantity $\eta$, our investigation revealed conflicting
results with respect to the relationship between observed number of
XRBs and cluster mass.  We performed a $\chi^2$ test on the slope in
$\eta$ for a variety of assumed star cluster age distributions and
found some cases where the slope was insignificant, while other cases
showed a distinctly positive slope.  No slope indicates the observed
number of XRBs per unit mass is independent of cluster mass, while a
positive slope in $\eta$ suggests more XRBs per unit mass are produced
in more massive clusters.  In the following discussion, we will
consider both a constant value in $\eta$ as well as a slope in $\eta$,
and the implications of each result.

Initially, we estimated cluster mass by fitting BC spectrophotometric
models to cluster $M_{K_s}$, assuming all clusters are $\sim$20 Myr.
Recognizing that this method depends on cluster age, we explored how
different assumptions of a cluster age distribution for the Antennae
affect $\eta$ and showed that $\eta$ varies by a factor of roughly
four; although including individual ages for the instant bursts case
will increase the variations in $\eta$ to a factor of 10.

We now proceed by comparing the mean value of $\eta$ for the four
different assumed age distributions, $\eta$ = 1.7$\times$10$^{-8}$
$M_{\sun}^{-1}$, to that predicted by models of young, massive
clusters.  We will compare $\eta$ to theoretical models discussed in
\citet{osk05} and \citet{sep05}.

In a recent study presented in \citet{osk05}, the author modeled X-ray
emission from young, massive star clusters, assuming a closed system
with constant mass, no dynamics and all stars are coeval, with cluster
metallicities of either z = 0.02 or z = 0.008.  These models predict
$\sim$2-5\% of all OB stars in a cluster should produce high mass
X-ray binaries (HMXBs).  In the models in \citet{osk05}, all clusters
are assumed to have masses of M$_{cl}$ = 10$^6$ $M_{\sun}$ with
stellar masses ranging from 1 -- 100 $M_{\sun}$.  Considering the
Salpeter initial mass function (IMF) of the form $\xi(M)$ =
$M_0M^{-2.35}$ and defining stars with masses $>8$ $M_{\sun}$ as ``OB
stars'', for our purposes here, we estimated 6\% of all stars in the
model clusters are OB stars.  Therefore, 1 -- 3$\times$10$^{-3}$ of
all stars in a cluster with an initial mass of 3$\times$10$^6$
$M_{\sun}$ (set by the Salpeter IMF) should produce an XRB.  Since the
Salpeter IMF implies there are 7$\times$10$^5$ stars in a cluster,
then these stars should produce 7 -- 22$\times$10$^2$ XRBs -- orders
of magnitude greater than the $\sim$49 observed in the Antennae.
Expressing $\eta$ as a fraction of XRBs-to-cluster mass, the models in
\citet{osk05} suggest $\eta$ ranges from 3--7$\times$10$^{-4}$
$M_{\sun}^{-1}$.  These values are greater by at least a factor of
1000 from our estimates for $\eta$.  Clearly, this predicts a much
larger number of compact object binaries than what we observed in the
Antennae.  \citet{osk05} note that they were unable to detect HMXBs in
three massive ($\sim10^4$ $M_{\sun}$) clusters which they predict
should contain between 1-3 HMXBs.  If our measured value for $\eta$
accurately describes the number of HMXBs formed, then it is not
surprising that \citet{osk05} fail to find any.  As pointed out by
these authors, future modeling of HMXB formation is needed to
understand the discrepancy between the predictions and observations of
XRB populations in starburst galaxies.

In another study, \citet{sep05} use the binary evolution and
population synthesis code, StarTrack \citep{bel02}, to investigate the
rate of XRB formation and ejection from young, massive clusters.  This
program tracks stellar parameters such as radius, luminosity, mass and
core mass.  The simulations are stopped at the formation of a compact
object.  The models include mass transfer in binaries and include
transient XRBs.  \citet{sep05} consider cluster masses ranging from
$5\times10^4$ $M_{\sun}$ to $5\times10^6$ $M_{\sun}$ and cluster ages
from 1 to $\sim$20 Myr and compute the average number of XRBs within
1--1000 pc of the cluster center.

Considering the typical cluster age in the Antennae is 20 Myr,
\citet{sep05} predict a $5\times10^4$ $M_{\sun}$ cluster should
contain 0.13 XRBs, while 15 XRBs should reside in a $5\times10^6$
$M_{\sun}$ cluster.  Here we assume that an XRB is associated with a
cluster if it is within 100 pc.  This separation is equivalent to
$1\farcs0$ at the distance of the Antennae (for $H_{0}$=75 km s$^{-1}$
Mpc$^{-1}$), which is our criteria for an XRB-cluster association
(Paper II).  These model predictions for XRB detections assume a
limiting X-ray luminosity of $L_X$ = 5$\times$10$^{35}$ ergs s$^{-1}$,
but the observed limiting luminosity in the Antennae is
2$\times$10$^{37}$ ergs s$^{-1}$.  Using the X-ray luminosity function
(XLF) for the Antennae defined in \citet{zez02c}, we scaled the XRB
results of \citet{sep05} to estimate what these models would predict
for the observed number of XRBs in the Antennae clusters.  Using a XLF
power law slope of $\alpha$ = -0.45 \citep{zez02c}, the models predict
0.02 XRBs are observed in a $5\times10^4$ $M_{\sun}$ cluster, while
2.7 XRBs should be seen in a $5\times10^6$ $M_{\sun}$ cluster, at the
luminosity limits of the X-ray observations.  Expressing these model
results as a fraction of XRBs-to-cluster mass, we can directly compare
them to our measured value for $\eta$ in the Antennae.  Doing so,
\citet{sep05} predict $\eta$ ranges from $4-5\times10^{-7}$
$M_{\sun}^{-1}$, with in a factor of five from our predictions for
$\eta$.  As mentioned by \citet{sep05}, several caveats exist for
their models including: 1) assumed binary fraction of unity which
could lead to over estimates of the mean number of XRBs per cluster,
2) the stellar, power-law IMF can affect the XRB fraction per cluster,
and 3) changes in the half-mass radius can strongly influence the
median XRB distance from the cluster.  These factors could potentially
explain the discrepancies between their models and our observations.

Since some forms of $\eta$ exhibit a distinctly positive slope, these
models may not always apply to $\eta$.  More importantly, a positive
slope has implications for star formation scenarios in clusters.  As
mentioned above, a positive slope implies more XRBs per unit mass are
produced in more massive clusters.  If an abnormally large number of
XRBs exist in a star cluster, then the progenitors of their compact
objects should also posses an over abundance.  Since the progenitors
are massive stars, this implies star formation in massive clusters
favors stars at the heavier end in mass.  This is not unusual.  Work
by \citet{sto05} suggest some of the largest clusters in the Milky Way
could have a top heavy mass function.

In this paper we introduced the quantity, $\eta$, relating the
fraction of X-ray sources per unit mass as a function of cluster mass.
Applying this function to the Antennae, we revealed several important
environmental implications for the X-ray sources in the Antennae.
Specifically, $\eta$ predicts a far different relationship between XRB
formation and cluster mass than that predicted by \citet{osk05} and is
broadly consistent with that predicted by \citet{sep05}.  Clearly,
future cluster modeling with particular emphasis on the relationship
between the number of XRBs in a galaxy and the galactic cluster
environment is essential to explain our current observations.
Furthermore, a $\chi^2$ test demonstrated the functional form of
$\eta$ did not always remain consistent with a single value, but for
some assumptions for a cluster age distributions, the slope had a
significantly positive value.  While this could imply a top heavy mass
function in massive clusters, our statistics are small.  We plan to
enlarge our statistical base by extending our observational study to
additional starburst galaxies.  We can then address whether the
properties of $\eta$ depend on an individual galaxy or are consistent
across all galactic environments.

\acknowledgments

The authors thank the staff of Palomar Observatory for their excellent
assistance in commissioning WIRC and obtaining these data.  WIRC was
made possible by support from the NSF (NSF-AST0328522), the Norris
Foundation, and Cornell University.  S.S.E. and D.M.C. are supported
in part by an NSF CAREER award (NSF-9983830).  We also thank
J.R. Houck for his support of the WIRC instrument project. The authors
are also grateful for many long and insightful discussions with
M.L. Edwards.

\vfill \eject

\clearpage

\begin{deluxetable}{lccc}
\tablecaption{$\Delta\Sigma\chi^2$ Statistics}
\tablewidth{0pt}
\startdata
\tableline
\tableline
Age Test & $\Delta\Sigma\chi^2$ & $\sigma_{\Delta\Sigma\chi^2}$ & Median $\Delta\Sigma\chi^2$ \\
\tableline
flux & 1.9 & --- & --- \\
20 Myr & 2.2 & --- & --- \\
100 Myr & 4.8$\times$10$^{-3}$ & --- & --- \\
Mengle & 1.0$\times$10$^{-2}$ & --- & --- \\
Instant & 0.86 & 1.7 & 0.10 \\
Uniform & 5.9 & 5.2 & 4.7 \\
Uniform\tablenotemark{1} & 5.8 & 5.3 & 4.7 \\
Power Law & 5.1 & 6.0 & 2.8 \\
Power Law\tablenotemark{1} & 4.3 & 6.6 & 2.1 \\
\enddata
\tablecomments{For instant, uniform and power law cases,
  $\Delta\Sigma\chi^2$ statistics are mean values.}
\tablenotetext{1}{Multiplied $\Delta\Sigma\chi^2$ by the sign of the
slope for the fitted line.}
\end{deluxetable}

\clearpage

\begin{figure}
\figurenum{1}
\plotone{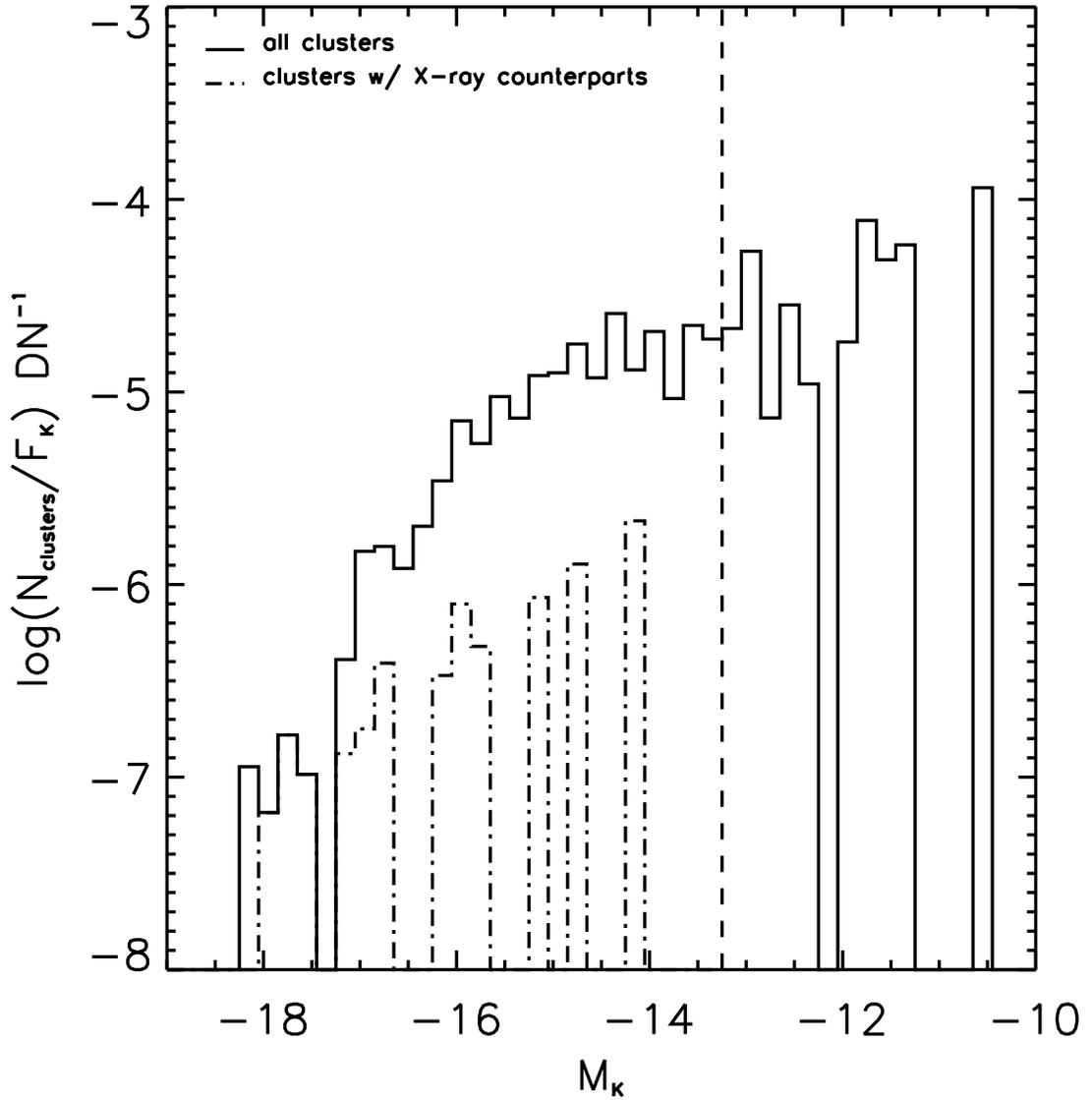}
\caption{Here we plot $M_{K_s}$ versus the number of clusters in each
  bin divided by mean flux in that bin.  Each bin is 0.2 mag.  Arguing
  that mass is proportional to flux, this graph shows the probability
  of finding a cluster with a given mass.  The dashed line signifies
  the magnitude cutoff, $M_{K_s} = -13.3$ mag.
\label{Fig.1}}
\end{figure}

\clearpage

\begin{figure}
\figurenum{2}
\plotone{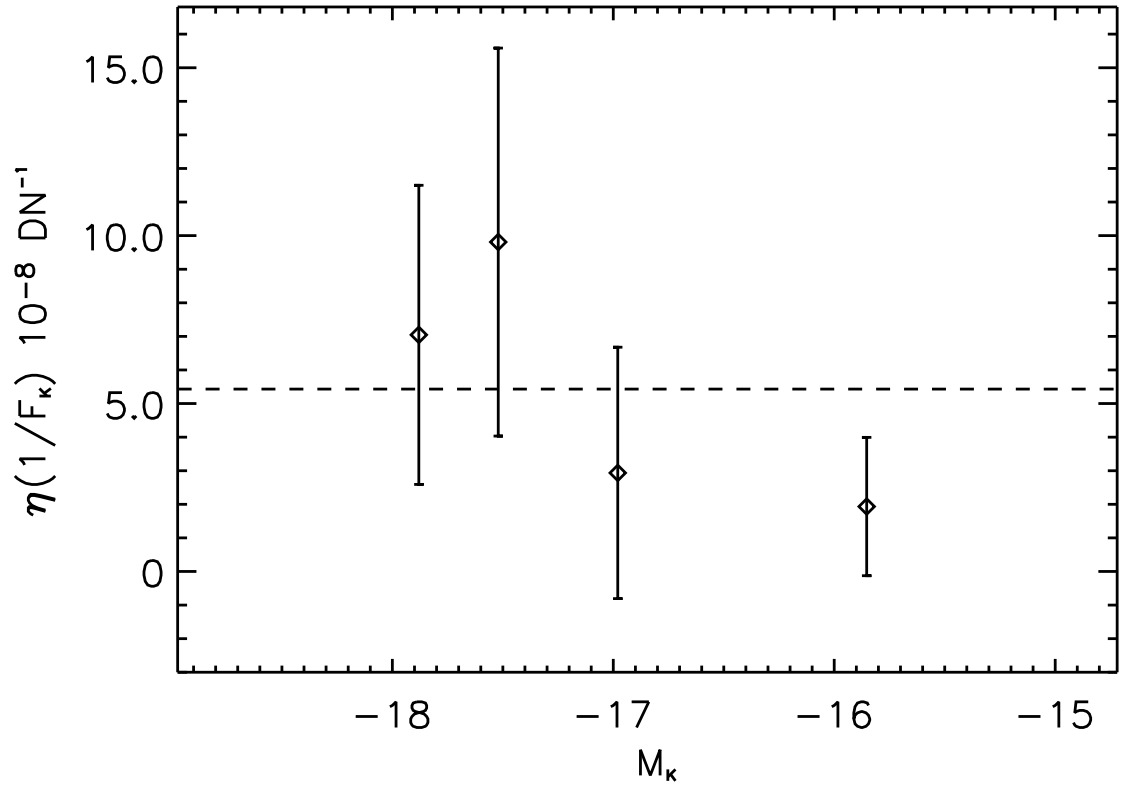}
\caption{This figure displays $\eta(F_{K_s})$ plotted versus
  $M_{K_s}$.  The bins are $F_{K_s}$ = $4\times10^6$ DN$^{-1}$ in
  size.  Error bars are the uncertainties in the mean value of $\eta$
  added in quadrature with the Poisson uncertainty in each bin.  The
  dotted line is the mean of the four $\eta(F_{K_s})$ values.
\label{Fig.2}}
\end{figure}

\clearpage

\begin{figure}
\figurenum{3}
\plotone{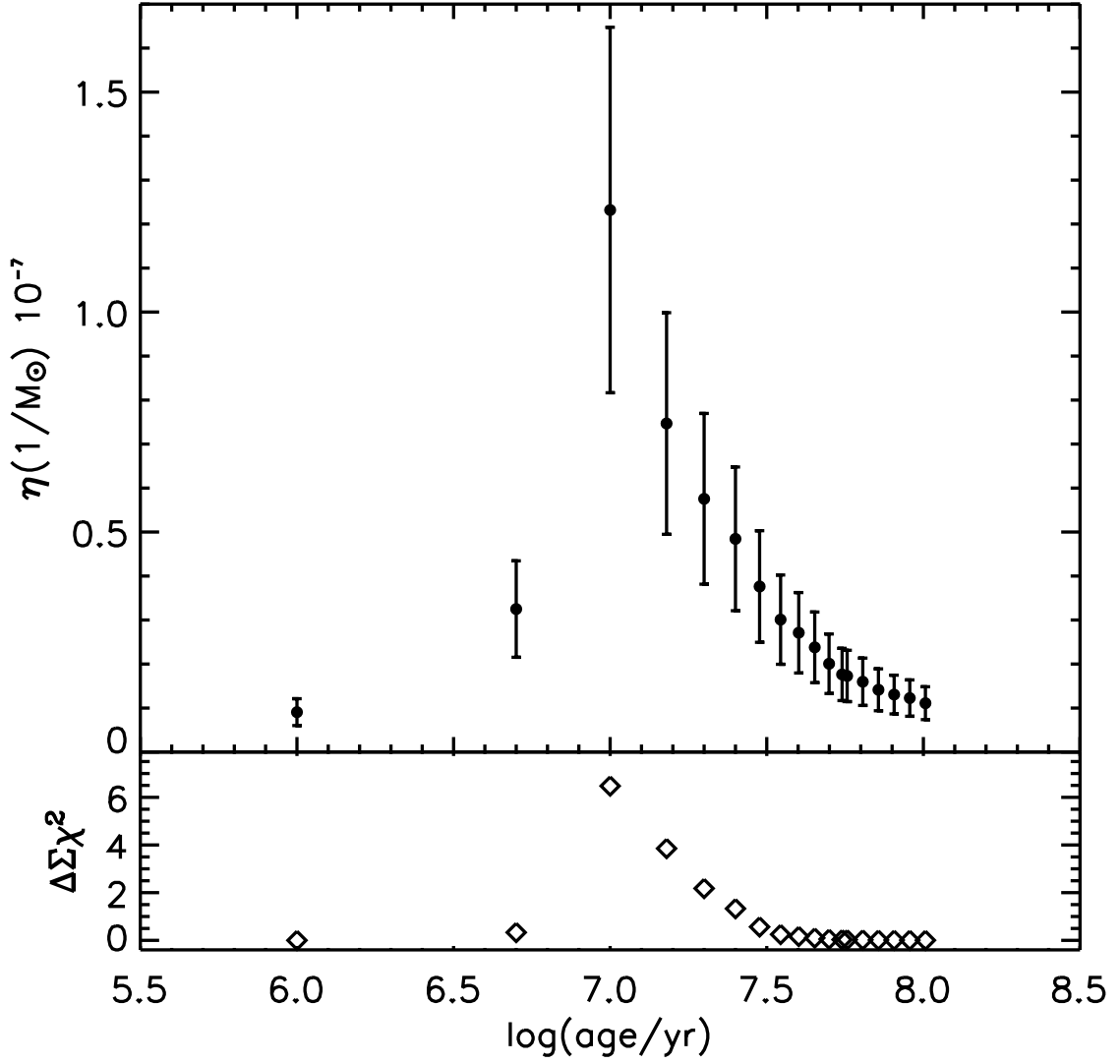}
\caption{Assuming an instant burst of star formation in the Antennae,
we plot the mean value of $\eta$ for a range in ages between 1 - 100
Myr.  Notice the factor of $\sim$10 range in $\eta$ as well as the
degeneracy in $\eta$ in this age range.  Error bars are uncertainties
in $\eta$.
\label{Fig.3}}
\end{figure}

\clearpage

\begin{figure}
\figurenum{4}
\plotone{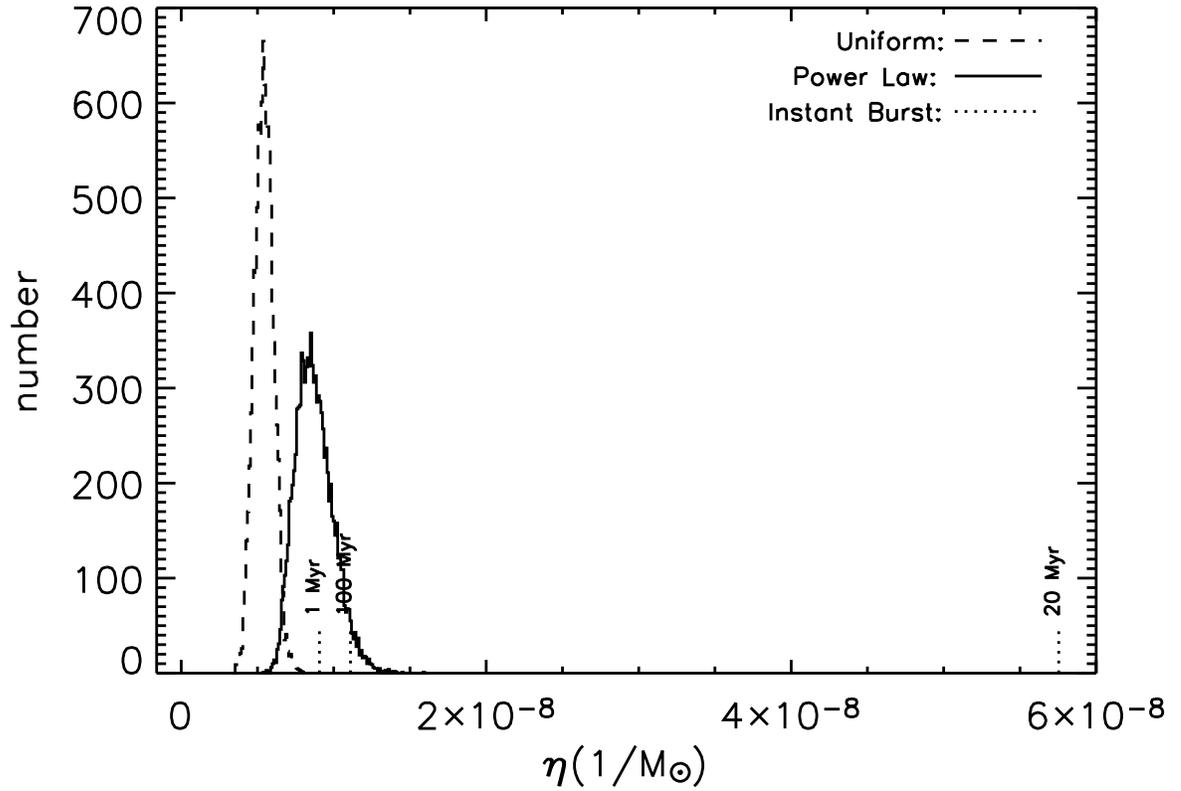}
\caption{Comparison between uniform and PL Monte Carlo simulations of
  $\eta$.  The peaks of each distribution vary by a factor of $\sim$2
  in $\eta$, indicating $\eta$ does not significantly change when we
  assume different age distributions for the Antennae.  We also plot
  the values of $\eta$ for instantaneous bursts at three different
  ages.
\label{Fig.4}}
\end{figure}

\clearpage

\begin{figure}
\figurenum{5}
\plotone{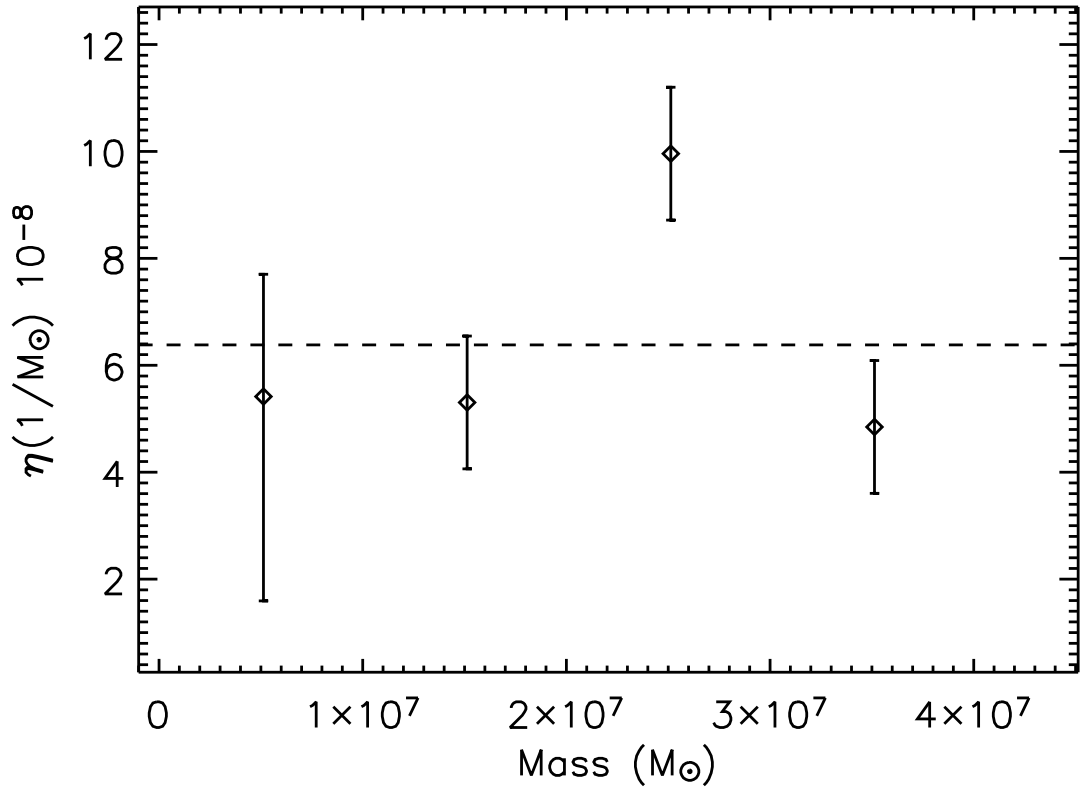}
\caption{Here $\eta$ is plotted versus cluster mass in units
  of $M_{\sun}$.  In this case, we computed cluster mass using ages
  provided by \citet{men05} (see \S3).  The bins are $M_{\sun}$ =
  1$\times10^7$ $M_{\sun}$ in size.  Error bars are the uncertainties
  in the mean value of $\eta$ added in quadrature with the Poisson
  uncertainty in each bin.  The dotted line is the mean value of the
  four $\eta(M_{\sun})$.
\label{Fig.5}}
\end{figure}

\clearpage

\begin{figure}
\figurenum{6}
\plotone{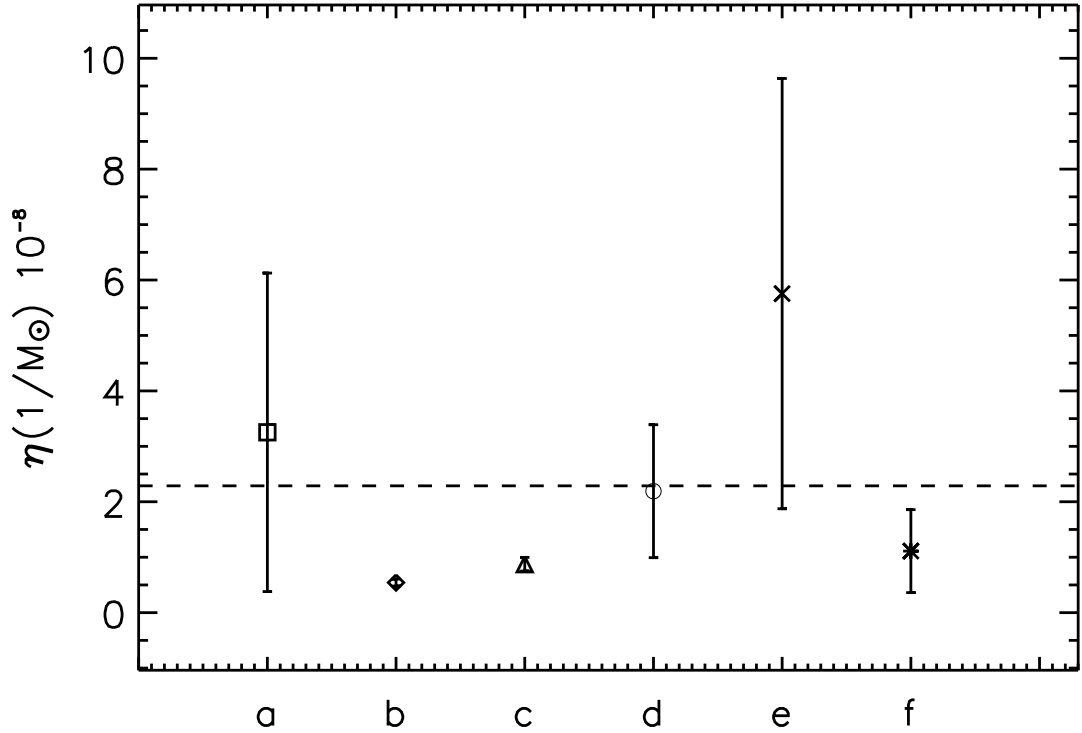}
\caption{Here we summarize how age affects $\eta$, assuming four
different age distributions for the Antennae clusters: instant burst
(a), uniform (b), power law (c) and derived ages by \citet{men05} (d).
Each value is the mean $\eta$ and includes 1-$\sigma$ error bars.  See
text for details.  Also included is $\eta$ for an instant starburst of
20 Myr (e) and 100 Myr (f).
\label{Fig.6}}
\end{figure}


\begin{thebibliography}{}

\bibitem[Anders et al.(2004)]{and04} Anders, P., de Grijs, R.,
Fritze-v. Alvensleben, U., \& Bissantz, N.  2004, \mnras, 347, 17

\bibitem[Belczynski et al.(2002)]{bel02} Belczynski, K., Kalogera, V., \&
Bulik, T.  2002, \apj, 572, 407

\bibitem[Brandl et al.(2005)]{bra05} Brandl, B.R., et al.  2005, \apj, 635, 280

\bibitem[Bruzual \& Charlot(2003)]{bru03} Bruzual, G. \& Charlot, S.  2003, 
	\mnras, 344, 1000

\bibitem[Cardelli, Clayton, \& Mathis(1989)]{car89} Cardelli, J.A., Clayton, 
	G.C., \& Mathis, J.S.  1989, \apj, 345, 245

\bibitem[Clark et al.(2005)]{cla05} Clark, D.M., et al. 2005, \apj,
  631, L109

\bibitem[Clark et al.(2007)]{cla07} Clark, D.~M., et al. 2007, \apj,
  658, 319

\bibitem[Fabbiano(1995)]{fab95} Fabbiano, G.  1995, in X-Ray
        Binaries, ed.  W.H.G. Lewin, J. van Paradijs, \& E.P.J. van den
        Heuvel (Cambridge: Cambridge Univ. Press), 390

\bibitem[Fabbiano et al.(1997)]{fab97} Fabbiano, G., Schweizer, F., \&
  Mackie, G.  1997, \apj, 478, 542

\bibitem[Fabbiano, Zezas, \& Murray(2001)]{fab01} Fabbiano, G., Zezas,
  A., \& Murray, S.S.  2001, \apj, 554, 1035

\bibitem[Fall(2006)]{fal06} Fall, S.M.  2006, \apj, in press

\bibitem[Fall, Chandar, \& Whitmore(2005)]{fal05} Fall, S.M., Chandar,
R., \& Whitmore, B.C.  2005, 631, L133

\bibitem[Gehrels(1986)]{geh86}  Gehrels, N.  1986, \apj, 303, 336

\bibitem[Harris(1996)]{har96} Harris, W.E.  1996, \aj, 112, 1487

\bibitem[Keeping(1962)]{kee62} Keeping, E.S.  1962, Introduction To
Statistical Inference, (Princeton; van Nostrano), p.202

\bibitem[Mengel(2005)]{men05} Mengel, S., Lehnert, M.D., Thatte, N.,
\& Genzel, R.  2005, \aa, 443, 41

\bibitem[Oskinova(2005)]{osk05} Oskinova, L.M.  2005, \mnras, 361, 679

\bibitem[Pooley et al.(2003)]{poo03} Pooley, D., et al.  2003, \apjl,
        591, L131

\bibitem[Portegies Zwart(2004)]{por04} Portegies Zwart, S.F., Hut,
	  P., McMillan, S.L.W., \& Makino, J.  2004, \mnras, 351, 473

\bibitem[Sepinsky, Kalogera, \& Belczynski(2005)]{sep05} Sepinsky, J.,
Kalogera, V., \& Belczynski, K.  2005, \apj, 621, L37

\bibitem[Stolte et al.(2005)]{sto05} Stolte, A., Brandner, W., Grebel,
  E.K., Lenzen, R., \& Lagrange, A.  2005, \apj, 628, L113.

\bibitem[Wilson et al.(2003)]{wil03} Wilson, J.C., et al.  2003,
Proc. SPIE, 4841, 451

\bibitem[Zezas et al.(2002a)]{zez02a}  Zezas, A., Fabbiano, G., Rots, A.H., \&
	Murray, S.S.  2002, \apj, 142, 239

\bibitem[Zezas et al.(2002b)]{zez02b}  Zezas, A., Fabbiano, G., Rots, A.H., \&
	Murray, S.S.  2002, \apj, 577, 710

\bibitem[Zezas \& Fabbiano (2002)]{zez02c} Zezas, A., \& Fabbiano, G.  2002,
\apj, 577, 726

\bibitem[Whitmore et al.(1999)]{whi99} Whitmore, B.C., Zhang, Q.,
        Leitherer, C., \& Fall, S.M.  1999, \aj, 118, 1551

\end{thebibliography}
\end{document}